\documentclass[dvips]{acta}
\usepackage{supertabular,lscape,epsfig}
\usepackage{amssymb}
\usepackage{amsmath}
\DeclareSymbolFont{ppa}{OT1}{ppl}{m}{it}
\DeclareMathSymbol{\vv}{\mathalpha}{ppa}{'166}

\newfont{\hb}{rphvb at 10pt}
\newfont{\hbo}{rphvbo at 10pt}
\newfont{\bitt}{rptmbi at 12pt}
\newfont{\bits}{rptmbi at 11pt}

\SetPages{1}{16}

\SetVol{60}{2010}

\begin{document}

\newcommand{\TabCapp}[2]{\begin{center}\parbox[t]{#1}{\centerline{
  \small {\spaceskip 2pt plus 1pt minus 1pt T a b l e}
  \refstepcounter{table}\thetable}
  \vskip2mm
  \centerline{\footnotesize #2}}
  \vskip3mm
\end{center}}

\newcommand{\TTabCap}[3]{\begin{center}\parbox[t]{#1}{\centerline{
  \small {\spaceskip 2pt plus 1pt minus 1pt T a b l e}
  \refstepcounter{table}\thetable}
  \vskip2mm
  \centerline{\footnotesize #2}
  \centerline{\footnotesize #3}}
  \vskip1mm
\end{center}}

\newcommand{\MakeTableSepp}[4]{\begin{table}[p]\TabCapp{#2}{#3}
  \begin{center} \TableFont \begin{tabular}{#1} #4 
  \end{tabular}\end{center}\end{table}}

\newcommand{\MakeTableee}[4]{\begin{table}[htb]\TabCapp{#2}{#3}
  \begin{center} \TableFont \begin{tabular}{#1} #4
  \end{tabular}\end{center}\end{table}}

\newcommand{\MakeTablee}[5]{\begin{table}[htb]\TTabCap{#2}{#3}{#4}
  \begin{center} \TableFont \begin{tabular}{#1} #5 
  \end{tabular}\end{center}\end{table}}

\newfont{\bb}{ptmbi8t at 12pt}
\newfont{\bbb}{cmbxti10}
\newfont{\bbbb}{cmbxti10 at 9pt}
\newcommand{\uprule}{\rule{0pt}{2.5ex}}
\newcommand{\douprule}{\rule[-2ex]{0pt}{4.5ex}}
\newcommand{\dorule}{\rule[-2ex]{0pt}{2ex}}
\newcommand{\totalno}{2786\ }
\newcommand{\totalgoodno}{1204\ }
\newcommand{\totalunno}{1490\ } 
\newcommand{\fHno}{67\ } 
\newcommand{\hHHno}{3\ } 
\newcommand{\hhHHHno}{3\ }
\newcommand{\hHHHno}{3\ }
\newcommand{\fHhhno}{two\ } 
\newcommand{\otherdoublem}{14\ } 
\newcommand{\doubleMwithCOMB}{19\ } 
\newcommand{\totalmultim}{92\ } 

\begin{Titlepage}
\Title{The Optical Gravitational Lensing Experiment.\\
The OGLE-III Catalog of Variable Stars.\\
VI.~Delta Scuti Stars in the Large Magellanic Cloud
\footnote{Based on observations obtained with the 1.3-m 
Warsaw telescope at the Las Campanas Observatory of the Carnegie
Institution of Washington.}}
\Author{R.~~P~o~l~e~s~k~i$^1$,~~
I.~~S~o~s~z~y~ñ~s~k~i$^1$,~~ 
A.~~U~d~a~l~s~k~i$^1$,~~ 
M.\,K.~~S~z~y~m~a~ñ~s~k~i$^1$,\\ 
M.~~K~u~b~i~a~k$^1$,~~ 
G.~~P~i~e~t~r~z~y~ñ~s~k~i$^{1,2}$,~~ 
£.~~W~y~r~z~y~k~o~w~s~k~i$^3$,\\
O.~~S~z~e~w~c~z~y~k$^2$~~ and~~ 
K.~~U~l~a~c~z~y~k$^1$}
{$^1$ Warsaw University Observatory, Al. Ujazdowskie 4, 00-478 Warszawa,
Poland\\ 
e-mail:
(rpoleski,soszynsk,udalski,msz,mk,pietrzyn,kulaczyk)@astrouw.edu.pl\\
$^2$ Universidad de Concepción, Departamento de Fisica, Casilla 160-C,
Concepción, Chile\\ 
e-mail: szewczyk@astro-udec.cl\\ 
$^3$ Institute of Astronomy, University of Cambridge, Madingley Road, Cambridge CB3 0HA, UK\\
e-mail: wyrzykow@ast.cam.ac.uk}
\vspace*{3pt}
\Received{December 15, 2009}
\end{Titlepage}
\vspace*{3pt}

\Abstract{The sixth part of the OGLE-III catalog of Variable Stars presents 
$\delta$~Sct pulsators in the Large Magellanic Cloud. Altogether \totalno
variable stars were found and amongst them \totalmultim are multi-mode objects,
including 67 stars pulsating in the fundamental mode and the first overtone
(F/1O), nine double-mode pulsators with various combinations of the first
three overtones excited (1O/2O, 2O/3O and 1O/3O pulsators), and two triple
mode (F/1O/2O) $\delta$~Sct stars. In total \totalunno of stars are marked
as uncertain, due to scattered photometry and small amplitudes. For
single-mode objects it was not possible to unambiguously identify pulsation
mode, however we suggest the most of the single-mode variable stars pulsate in
the first overtone. }{Catalogs -- $\delta$~Scuti -- Stars: oscillations
-- Magellanic Clouds}

\vspace*{3pt}
\Section{Introduction} 
$\delta$ Scuti stars are main sequence (MS) or early post-MS pulsators
occupying lower part of the classical instability strip. It is much harder
to discover these stars at the same distance as other pulsators lying in
the classical instability strip, Cepheids and RR~Lyr stars, because
generally they are fainter and their light variations have smaller
amplitudes. Until now only very few pulsators of this type were known in
the galaxies other than the Milky Way.

Mateo \etal (1998) detected 20 $\delta$~Sct stars in the Carina dwarf
spheroidal galaxy and those were the first extragalactic stars of this
type. Their short periods and position in the color--magnitude diagram
(CMD) suggest they belong to a subgroup of $\delta$~Sct variable stars --
SX~Phe stars. Members of this subgroup are Population II objects and are
frequently identified as blue stragglers in the globular clusters. Kaluzny
and Rucinski (2003) searched for short-period variable stars in the field
of the Large Magellanic Cloud (LMC) open cluster LW 55 and revealed eight
objects.  Soszyñski \etal (2003) searched for RR~Lyr stars in the LMC
using data obtained by the second phase of the Optical Gravitational
Lensing Experiment (OGLE). As a by-product a list of 37 short-period
variable stars was presented. McNamara \etal (2007) classified 24 of these
stars as LMC $\delta$~Sct stars. Di~Fabrizio \etal (2005) found one
$\delta$~Sct star in the LMC. Huber \etal (2005) announced a discovery of a
large number of $\delta$~Sct stars in the LMC by the SuperMACHO project,
but no further details were published.  Kaluzny \etal (2006) found one
$\delta$~Sct pulsator in the LMC disk field. Around 100 other $\delta$~Sct
or SX~Phe stars were found in NGC~6822 irregular galaxy and also Fornax,
Coma and Leo IV dwarf spheroidal galaxies (Baldacci \etal 2005, Poretti
\etal 2008, Greco \etal 2009, Musella \etal 2009, Moretti \etal 2009).

The OGLE-III project collected photometric data for about 32 million stars
in the LMC for eight years. Typically 500 {\it I}-band and 50 {\it V}-band
images of each field were taken during that time. Although the original
purpose of the project was to find the gravitational microlensing events
the resulting data base allows discoveries of unprecedented number of
variable stars. This paper is the sixth part of the OGLE-III catalog of
Variable Stars (OIII-CVS) and presents
\totalno $\delta$~Sct stars observed in the direction of the LMC. 
Out of them \fHno objects pulsate simultaneously in the fundamental mode
and the first overtone. The distance to the LMC is known with an
uncertainty of $\approx 3\%$ (Pietrzyñski \etal 2009) and gives additional
constrain on absolute luminosity for modeling of multi-mode pulsators
presented in this paper.

Current observational works on $\delta$~Sct stars are rather focused on
multi-site campaigns investigating individual field objects. They allow
discoveries of many pulsation modes in observed stars. Good example is the
campaign carried out by Breger \etal (2005) on FG~Vir. The catalog prepared
on the basis of the OGLE data gives less detailed information about
individual stars but covers much larger number of objects.

Section~2 describes observational data and their reduction.  Section~3
gives detail information on our selection procedure. The catalog itself is
described in Section~4. The discussion of discovered variable stars is
given in Section~5. We end with conclusions in Section~6.

\Section{Observations and Data Reduction}
The OGLE-III photometric data had been collected between 2001 and 2009 with
the 1.3-m Warsaw telescope situated at Las Campanas Observatory, Chile. The
observatory is operated by the Carnegie Institution of Washington. The
telescope was equipped with the ``second generation'' camera containing
eight SITe $2048\times4096$ CCD detectors. Pixel size of 15~$\mu$m
corresponds to 0.26 arcsec/pixel scale. Total field of view was around $35
\times35.5$ arcmin. The readout noise varied between different detectors
from 6~e$^-$ to 9~e$^-$ and gain was set to about 1.3~e$^-$/ADU. Details of
the instrumental setup were described by Udalski (2003).

The photometry was obtained using the Difference Image Analysis (DIA)
me\-thod (Alard and Lupton 1998, Alard 2000, Wo¼niak 2000). The final data
reduction procedure was described in detail by Udalski \etal (2008a). The
116 OGLE-III fields in the LMC cover an area of almost 40 square degrees
and around 32 million of stars. For central 4.5 square degrees of the LMC
the OGLE-II photometry is available (Szymañski 2005). Whenever it was
possible we connected the OGLE-III photometric data with the OGLE-II
ones. To assure these additional data are in the same photometric system we
added to the OGLE-II magnitudes the difference between mean brightnesses in
OGLE-III and OGLE-II data. The objects lying near the edges of the CCD
chips are in some cases present in two or even three fields because the
OGLE-III fields slightly overlap. We combined the photometry from the
adjacent fields. For the particular objects, especially very faint ones,
either cross identification or brightness transformation may be
defective. No interstellar extinction corrections were applied.

The photometric uncertainties of the DIA photometry are known to be
underestimated. To make them more reliable we have corrected them using
formula $\sigma_{\rm new}=\sqrt{\left(\gamma\,
\sigma_{\rm old}\right)^2+\epsilon^2}$. The coefficients $\gamma$ and
$\epsilon$ differ for the different fields and chips and were calculated by
J.~Skowron (private communication). The details are described by
Wyrzykowski \etal (2009).

\Section{Selection Criteria}
We expected most of the $\delta$~Sct stars in the LMC would have {\it
I}-band magnitudes in the range between 19~mag and 21~mag. It is close to
the detection limit of the OGLE-III photometry and the standard deviation
of magnitudes rises in this interval from 0.05~mag to 0.3~mag (see Udalski
\etal 2008, Fig.~2 and 6). Thus, selection of candidates was performed in
a few steps during which we tried to remove interfering variable stars,
such as short period Cepheids, RR~Lyr stars, $\beta$~Cep stars and
eclipsing variables, as well as photometric artifacts which are more of a
problem for fainter stars, but not to remove the most sound $\delta$~Sct
stars.

\subsection{Selection of Single-Mode Pulsators}
At the beginning of the investigation of the OIII-CVS we performed a
massive period search. For each star observed in the LMC the frequency
range from 0~1/d to 24~1/d with a step of 0.00005~1/d was inspected using
the {\sc FNPeaks} code (Z.~Ko³aczkowski, private communication). To search
for $\delta$~Sct stars we selected only the stars with the most prominent
peak in the periodogram corresponding to the period shorter than 0.249~d
and with the signal to noise ratio higher than 5 (if not stated otherwise
signal to noise ratio -- S/N -- mentioned below corresponds to a value from
these periodograms). We note here 0.24~d period was taken as a boundary
between first overtone Cepheids and $\delta$~Sct stars by Soszyñski \etal
(2008). In the present study we did not find any good candidate for the
single-mode pulsator with period longer than 0.24~d.

The period distribution of the stars selected in such a way contained two
prominent peaks near 1/5~d and 1/6~d, thus almost all the stars with
periods close to these values were left out. Visual inspection of the light
curves ensured us that the majority of these stars were indeed
artifacts. The DIA photometry is influenced by variability of nearby stars
thus from the groups of objects lying close to each other on the sky and
possessing similar periods only one object was selected and all the
remaining were removed. Using the first two parts of the OIII-CVS
(Soszyñski \etal 2008, 2009) we have also removed stars lying near the
previously found Cepheids and RR~Lyr type stars.

For further selection and analysis, outlying photometric points as well as
ones with the highest values of the magnitude uncertainties were
removed. For all objects we tried to correct the period found during the
massive period search using the method described by Schwarzenberg-Czerny
(1996). For most of the stars the difference between two period estimations
was very small. Some periods were changed to their $\pm~1/\mathrm{d}$
harmonics and for a group of objects we could not confirm initially derived
periods and these objects were left out.

The next step of the selection was the visual inspection of the remaining
candidates. Stars with large amplitudes were marked as eclipsing binaries,
if folded with period twice longer than estimated above they showed
different depths of minimum light or different shapes of two halves of the
light curve. In this stage some objects were classified as artifacts.

For more sophisticated elimination of the eclipsing systems we have fitted
all the {\it I}-band light curves with the Fourier series (Simon and Lee
1981) consisting of two to five terms depending on the quality of the
light curve. The number of the {\it V}-band measurements in the OGLE-III
photometry is much smaller than in the {\it I}-band. Instead of fitting the
Fourier series to the {\it V}-band data we decided to transform the fitted
{\it I}-band light curves. Using the Levenberg-Marquardt algorithm (Press
\etal 1992) for each star we have found three coefficients -- phase
difference, difference in the mean brightness and amplitude ratio -- that
minimize the $\chi^2$ between {\it V}-band data and the transformed fitted
{\it I}-band light curve. We found for most of the obvious eclipsing
systems that the amplitudes in each of the pass bands are nearly
equal. This should be true for contact systems as temperatures of
components are similar. For pulsating stars the effective temperature
changes during each cycle and thus amplitude is different in different pass
bands. In many cases our candidates had very noisy light curves thus the
comparison of amplitudes was performed only for objects with ${\rm
S/N}>7.5$. Fig.~1 shows a plot of the ratio of amplitude of the first
Fourier terms in {\it V} ($A_1(V)$) and {\it I} ($A_1(I)$). One can see
that points in Fig.~1 tend to group around two regions. This is more
evident if higher S/N limit is assumed. For the first group (eclipsing
systems) $A_1(V)\approx A_1(I)$ and for the second (pulsating stars)
$A_1(V)\approx1.5 A_1(I)$. The line which best separates the two groups is
in our opinion $A_1(V)=1.316 A_1(I)$ and it was adopted as a border line
between two groups. In obvious examples (detached eclipsing binaries and
double mode pulsators) we did not take into account the amplitude ratio.
\begin{figure}[htb]
\centerline{\includegraphics[width=11.5cm]{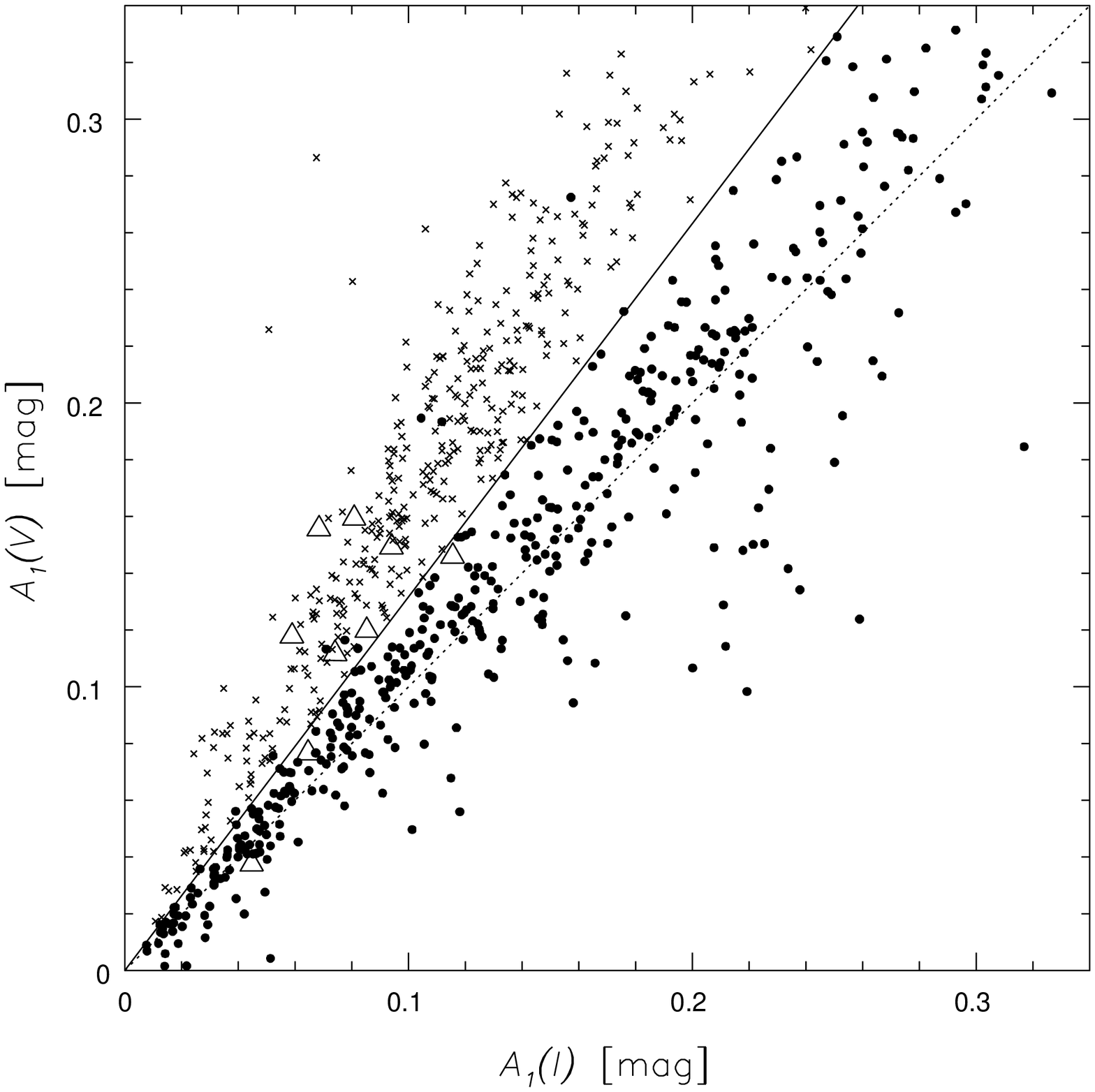}}
\FigCap{Amplitudes of the first terms in Fourier series in {\it I}-band ($A_1(I)$) 
and {\it V}-band ($A_1(V)$) data. Stars classified as eclipsing binaries
are marked by filled circles. Crosses mark single mode $\delta$~Sct stars
and open triangles -- F/1O ones. Solid line represents the adopted
boundary between eclipsing binaries and pulsating stars.  Dashed line
represents $A_1(V)=A_1(I)$ relation.}
\end{figure}

We classified 84 stars from our list as candidates for $\beta$~Cep. First
such stars in the LMC were discovered by Pigulski and Ko³aczkowski
(2002). The more detailed search, probably with a smaller limit on the S/N,
and the analysis of these pulsators will be done in one of the next parts
of the OIII-CVS.

Many candidates for short-period pulsating stars were found among the red
clump and red giant stars. Some of these objects may be $\delta$~Sct
pulsators blended with other star but most of these stars were spurious
candidates. In order to clean candidate list from artifacts situated in
this region of the CMD we have removed objects with the ${\rm S/N}<
6.2$ and with $\phi_{21}=\phi_2-2\phi_1$ smaller than 3 or greater than 5.
It is shown in Section~4 that most of the $\delta$~Sct stars have light curves
described by $3<\phi_{21}<5$. After applying this procedure the red giant
branch and the red clump are barely seen on the CMD of candidates.

The next step of our cleaning procedure was removing numerous eclipsing
systems amongst stars with ${\rm S/N}<7.5$. These stars had periods near
our selection limit (\ie half of the orbital period was close to 0.24~d),
were typically 1.5~mag fainter in the {\it I}-band than the $\delta$~Sct
stars with similar periods and had $\phi_{21}$ values close to 0, contrary
to $\phi_{21}\approx4$ for the $\delta$~Sct stars. All the objects lying
inside the selected region on $\log{P}$--{\it I} magnitude diagram were
removed from the list of candidates.

In order to identify galactic objects in the foreground of the LMC we
estimated for each candidate the proper motion using profile photometry of
the OGLE-III data (Udalski \etal 2008a) which also gives astrometric
position of a star for each observation. Since we only wanted to select
high proper motion objects in our candidate list we did not correct
measured positions for parallax or differential refraction effects. We left
out a few dozens of mostly red high proper motion stars. These objects are
nearby stars for which observed light variations are most probably caused
by rotation.

Finally, we added to the list of $\delta$~Sct stars altogether eleven stars
with periods longer than adopted limit (0.249~d). One of them
(OGLE-LMC-DSCT-0765 = LMC\_SC11\_241461) came from Soszyñski \etal (2003)
and all other stars were found during the earlier searches for variable
stars in the LMC (Soszyñski \etal 2008, 2009). The catalog contains also
five stars for which the mean luminosities vary in our data.

Finding the clear cut between variable stars and artifacts is especially
difficult task for the noisy light curves like the ones analyzed in the
present study. We decided to remove single-mode candidates with ${\rm
S/N}<5.2$ and assign objects with $5.2<{\rm S/N}<6$ as uncertain in our
catalog. Fig.~2 presents sample light curves of stars with ${\rm S/N} < 6$
(left panels) and ${\rm S/N}>6$ (right panels). We note that even though
our classification may be wrong for individual stars, the catalog contains
mostly true pulsating variable stars and may be used for statistical
investigations.
\begin{figure}[htb]
\vglue-3mm
\centerline{\includegraphics[height=12cm, angle=270]{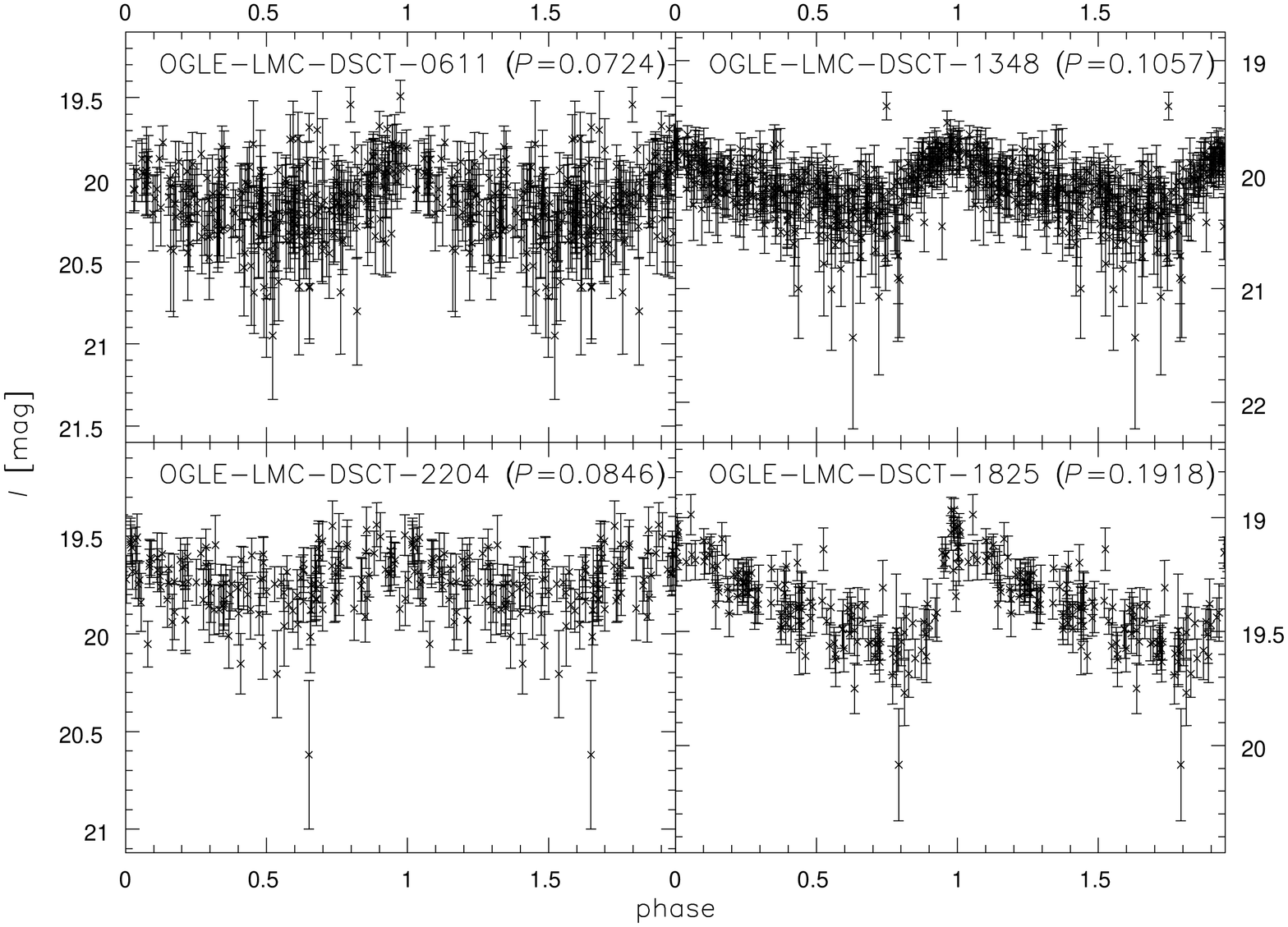}}
\FigCap{Sample light curves of $\delta$~Sct stars marked as 
uncertain ({\it left panels}) and not marked ({\it right panels}). In each
panel OGLE designation and period are given. For clarity randomly chosen
1/3 of points is shown.}
\end{figure}

\subsection{Selection of Multi-Mode Pulsators}
The selection of candidates for multi-mode pulsators was performed in two
ways. In the first one the periodograms from the massive period search were
inspected for prominent secondary peaks with ${\rm S/N}>5$ and not caused
by the daily aliases. The second method was the prewhitening of all
previously selected $\delta$~Sct candidates for which we derived the
periods with the Schwarzenberg-Czerny (1996) method. After the
prewhitening, the periodograms were calculated once more using the {\sc
FNPeaks} software and searched for peaks with ${\rm S/N}>5$ (for
combination frequencies the limit was lowered to 4). In both methods we
also checked several stars with the peaks ${\rm S/N}<5$ but closer
examination revealed that all of these peaks (except one with ${\rm S/
N}=4.92$) were spurious results.

The Petersen diagram (\ie the plot of logarithm of the longer period $P_L$ 
\vs ratio of the shorter to the longer period $P_S/P_L$) revealed that 
most of the double mode pulsators oscillate in the fundamental mode (F) and
the first overtone (1O). For $\delta$~Sct stars the period ratios of these
stars lie in the range 0.75--0.79 what is predicted by models (Petersen and
Christensen-Dalsgaard 1996) and confirmed by observations (Alcock \etal
2000, Pigulski \etal 2006). We used the {\sc Period04} software (Lenz and
Breger 2005) for the closer examination of all multi-mode
candidates. Finally, we found \totalmultim multi-mode pulsators. The
identification of 1O/2O, 2O/3O and 1O/2O period ratios are based on Olech
\etal (2005, their Fig.~6). Fig.~3 presents the Petersen diagram for our
multi-mode $\delta$~Sct stars.
\begin{figure}[htb]
\centerline{\includegraphics[width=11.5cm]{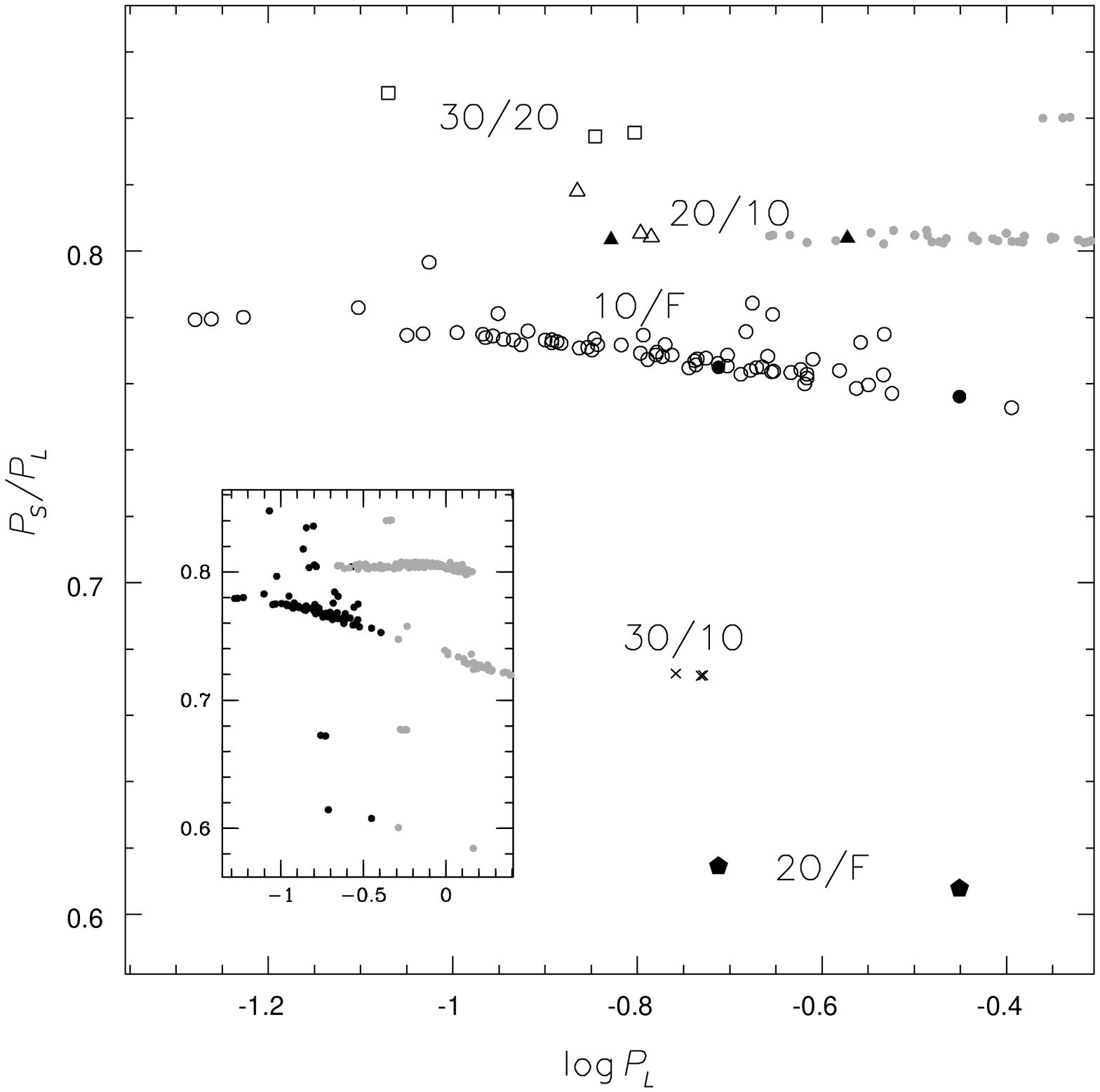}}
\FigCap{Petersen diagram for multi-mode pulsators with mode identified. 
Open symbols correspond to the objects with two modes identified and filled
black ones -- to objects with three modes identified (three symbols per
star). Circles represent F/1O modes, triangles -- 1O/2O, squares -- 2O/3O,
pentagons -- F/2O and crosses -- 1O/3O (three double mode objects, two of
them the very near each other). Gray dots show OGLE-III Cepheids (Soszyñski
2008). The insert shows $\delta$~Sct stars (black dots) and Cepheids (gray
dots).}
\end{figure}

We did not find any candidates for $\delta$~Sct stars in eclipsing binary
systems.

\Section{Catalog}
Our catalog of $\delta$~Sct stars in the LMC includes \totalno variable stars,
of which \totalunno are flagged as uncertain. There are \totalmultim
objects with more than one mode detected. This catalog is much larger than
the biggest catalog of $\delta$~Sct stars published so far (Rodr\'iguez
\etal 2000) which contained around 600 stars in the Galaxy.
    
This part of the OIII-CVS is available in the electronic form only from the
OGLE Internet archive:
\begin{center}
{\it 
http://ogle.astrouw.edu.pl/\\
ftp://ftp.astrouw.edu.pl/ogle/ogle3/OIII-CVS/lmc/dsct/
}
\end{center}
The {\sc ftp} site contains a full list of the identified $\delta$~Sct stars in
the {\sf ident.dat} file. All the stars are sorted according to their right
ascension and have identifiers in the form of OGLE-LMC-DSCT-NNNN, where
NNNN is a consecutive four digit number. The columns contain: OIII-CVS
identifier, the OGLE-III photometric map identifier (the subfield and the
number from Udalski \etal 2008b), our classification (SINGLEMODE or
MULTIMODE), RA and DEC coordinates for epoch 2000.0 and OGLE-II designation
(Udalski \etal 2000) for stars in the central region of the LMC.

Basic parameters of the single-mode pulsators are stored in the file {\sf
singlemode.dat} which contains: OIII-CVS identifier, intensity mean {\it I}
and {\it V}-band magnitudes, period and its uncertainty, the {\it I}-band
amplitude peak-to-peak, and two Fourier coefficients of the light curve --
$\phi_{21}$ and $R_{21}$ ($R_{21}=A_2/A_1$ where $A_1$ and $A_2$ are the
amplitudes of the first two Fourier series terms). File {\sf multimode.dat}
contains basic parameters for multi-mode $\delta$~Sct stars. The number of
frequencies found (up to eight) forced us to arrange this file
differently. For each star the number of rows is equal to the number of
frequencies detected. The first column contains OIII-CVS identifier only in
the first row for a given star. Other rows have ``--'' mark in that
column. The rest of the row contains: period and its error, time of the
maximum light, amplitude in magnitudes and remarks such as mode
identification or combination frequency identification at the end.

The file {\sf remarks.dat} contains additional information on our
$\delta$~Sct stars. Most records show uncertain objects. For nineteen stars
significant proper motion was found and its value is given in each
case. Since we were able to determine proper motions for Galactic stars the
objects with given proper motion are $\delta$~Sct stars belonging to the
Milky Way.

\Section{Discussion}
\subsection{Multi-Mode Pulsators}
We found \fHhhno triple mode pulsators pulsating simultaneously in the
fundamental mode and the first two overtones (F/1O/2O). Among stars with
two radial modes identified there are
\fHno F/1O, three 1O/2O, three 2O/3O and three 1O/3O pulsators.
Out of them we found combination frequencies for \doubleMwithCOMB objects. 
Fundamental mode periods are longer than assumed selection limit 
(0.249 d) for eight F/1O and one F/1O/2O pulsator. 

In Fig.~3 there are a few F/1O pulsators with period ratios higher by
0.01--0.02 than for the remaining stars and no pulsators with evidently
smaller ratios. We note that for previously discovered F/1O $\delta$~Sct
stars there are only a couple of stars with the period ratio smaller than
typical and several with period ratio higher than typical (Pigulski \etal
2006).  In our sample there is also one 1O/2O object with $P_S/P_L$ ratio
higher by 0.012 than for the remaining 1O/2O objects. Its classification is
less clear. Some of both the F/1O and 1O/2O objects mentioned above may be
nonradial pulsators for which it is not possible to photometrically
identify pulsation mode. The mode identification for the star shown in
Fig.~3 with $\log P_L\approx-1$ and $P_S/P_L\approx0.8$ is less clear and
this may be either F/1O or 1O/2O pulsator.

The insert in Fig.~3 shows that the period ratios for double mode
$\delta$~Sct stars and Cepheids, with the same modes excited, follow the
same relations. As it was mentioned before, the observational boundaries
between these two groups of stars are a convention. Theoretical discussion
of evolutionary status of the short period Cepheids was done by Baraffe
\etal (1998) and Dziembowski and Smolec (2009).

The highest number of frequencies found in one object is eight
(OGLE-LMC-DSCT-0048). Except for the fundamental mode frequency ($f_{\rm
F}$) and the first overtone one ($f_{\rm 1O}$) also six their combinations
were found: $2f_{\rm F}$, $f_{\rm F}+f_{\rm 1O}$, $f_{\rm 1O}-f_{\rm F}$,
$3f_{\rm F}$, $2f_{\rm F}+f_{\rm 1O}$ and $2f_{\rm F}-f_{\rm 1O}$.

The period--luminosity relations for radial modes identified in multi-mode
objects are shown in Fig.~4 separately in the {\it V}- and {\it I}-bands as
well as for the reddening free Wesenheit index ($W_I=I-1.55(V-I)$).
\begin{figure}[htb]
\vglue-3mm
\centerline{\includegraphics[width=11.7cm]{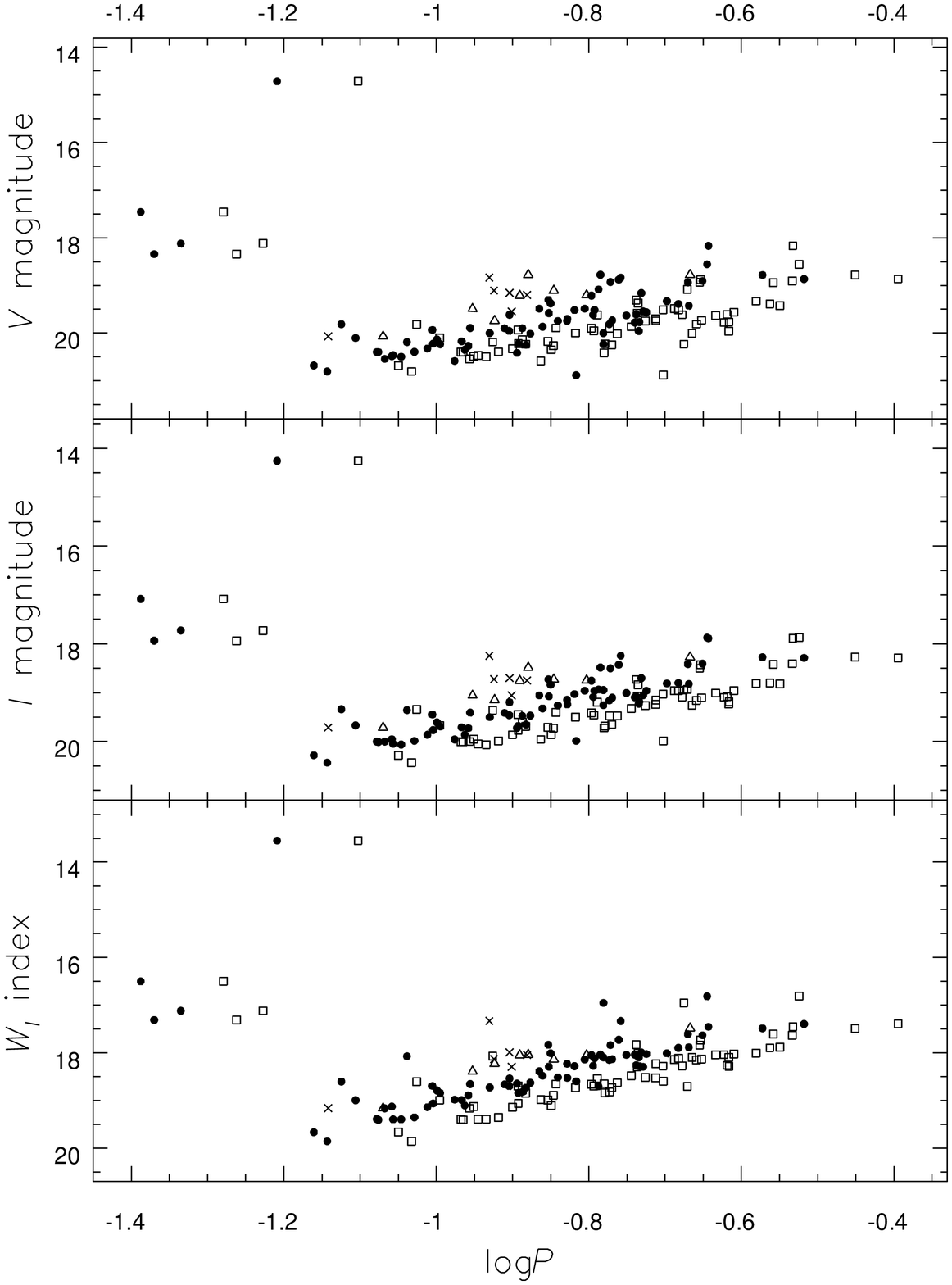}}
\vspace*{-3pt}
\FigCap{Period--luminosity diagrams for multi-mode $\delta$~Sct stars with 
the radial modes identified. Open squares, filled circles, open triangles
and crosses correspond to fundamental mode and first, second and third
overtone, respectively.}
\end{figure}
The linear fits were found by the least squares method with $3\sigma$
clipping. The following relations were derived for fundamental modes:
\begin{eqnarray*}
V  &=&-2.84(\pm0.26)\log{P}+17.68(\pm0.20)\qquad\sigma=0.30~{\rm mag} \\
I  &=&-3.20(\pm0.24)\log{P}+16.88(\pm0.19)\qquad\sigma=0.30~{\rm mag} \\
W_I&=&-3.68(\pm0.17)\log{P}+15.76(\pm0.13)\qquad\sigma=0.20~{\rm mag} \\
\end{eqnarray*}
\vskip-5mm
and for the first overtone modes:
\begin{eqnarray*}
V  &=&-3.17(\pm0.28)\log{P}+17.03(\pm0.25)\qquad\sigma=0.34~{\rm mag} \\
I  &=&-3.35(\pm0.25)\log{P}+16.36(\pm0.22)\qquad\sigma=0.30~{\rm mag} \\
W_I&=&-3.74(\pm0.17)\log{P}+15.28(\pm0.15)\qquad\sigma=0.19~{\rm mag} \\
\end{eqnarray*}
\vskip-3mm
The magnitudes were not corrected for interstellar extinction. These
relations predict similar brightnesses as for classical Cepheids (Soszyñski
\etal 2008) in the period range of 0.22--0.249 d in which, according to
assumed boundaries, $\delta$~Sct and Cepheids meet. Because of lower
brightnesses and smaller number of objects the uncertainties of the fitted
parameters are an order of magnitude higher than for Cepheids. Our
$\log P$--{\it V} relation is consistent within the uncertainties with
earlier studies (McNamara \etal 2007).

One can easily notice on the period-luminosity relations four F/1O objects
with short periods which are evidently brighter than the remaining F/1O
stars. Among these stars there are two for which we found significant
proper motions: OGLE-LMC-DSCT-0048 and OGLE-LMC-DSCT-0972. These two stars
are clearly located in the Galaxy.

The catalog also contains \otherdoublem multi-mode pulsators with no modes
identified. For six (\ie 43\%) of these stars we found more than two
frequencies while 27\% of objects with identified radial modes possess more
than two frequencies. Only for two of these multi-mode pulsators we found
combination frequencies. The ratios of periods are close to one. The
exception is a star OGLE-LMC-DSCT-1684 for which we found period ratio
equal to 0.309. If both periods correspond to radial modes then the shorter
one is a very high order mode and this is an unusual object. For five
multi-mode pulsators without mode identification we found significant
proper motions what means that these objects are Galactic foreground
$\delta$~Sct stars.

\subsection{Single-Mode Pulsators}
The period-luminosity relations for single-mode $\delta$~Sct stars are
shown in Fig.~5.
\begin{figure}[htb]
\centerline{\includegraphics[width=12.5cm]{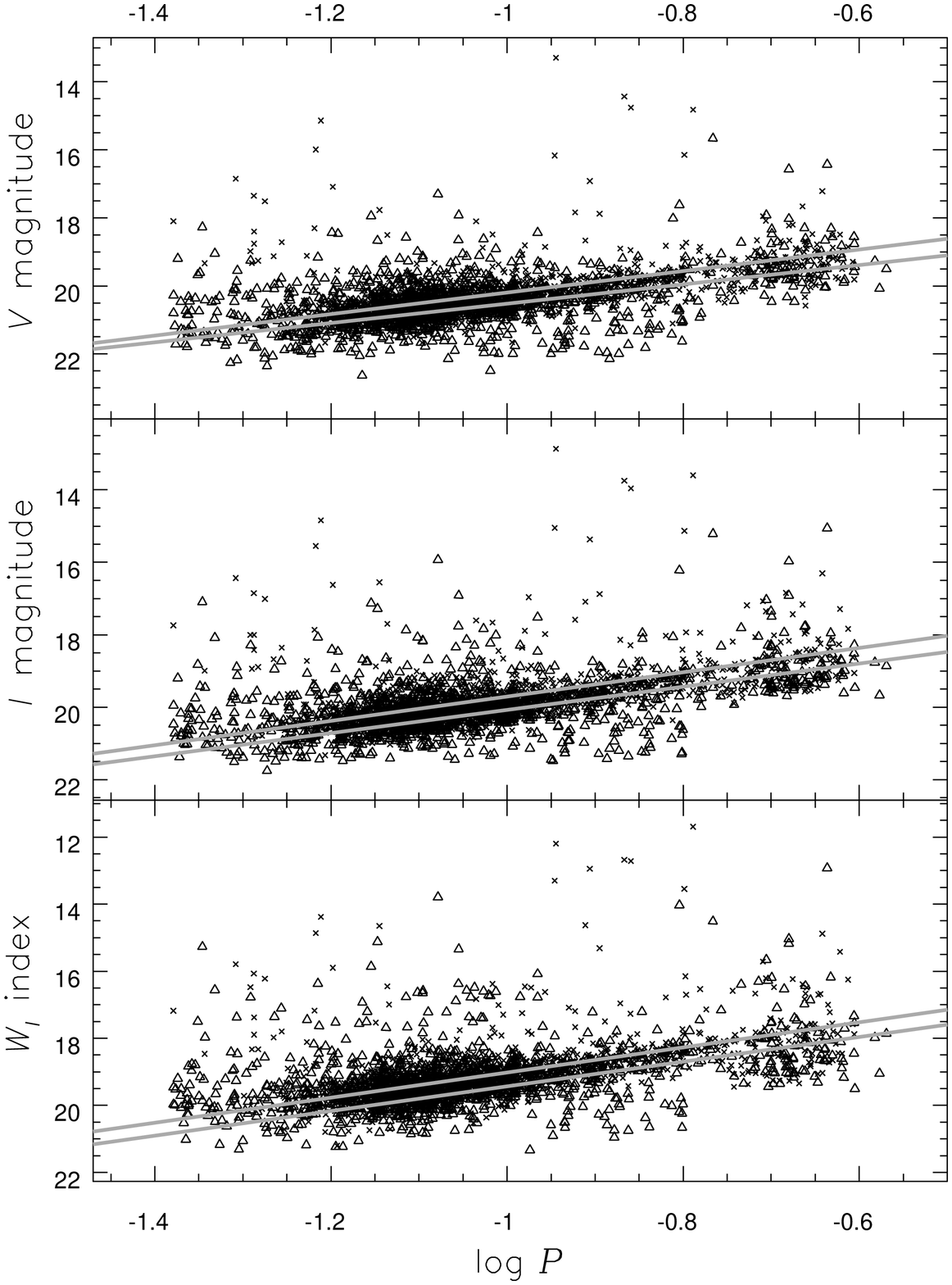}}
\FigCap{Period-luminosity relations for single-mode $\delta$~Sct 
pulsators. Crosses correspond to the objects marked as uncertain while all
the other single-mode $\delta$~Sct stars are represented by open
triangles. The gray lines in each panel show linear fits for multi-mode
objects. The bottom line in each panel corresponds to the fundamental mode
pulsators and the top one -- to the first overtone pulsators.}
\end{figure}

The gray lines show linear fits obtained in Section~4 for multi-mode
pulsators. It is not possible to pinpoint pulsation mode for each
individual object. As most of the stars are located symmetrically on the
two sides of the line corresponding to the 1O modes we suspect most of the
single-mode pulsators to be the 1O ones.

The group of $\delta$~Sct stars is typically divided into high and low
amplitude objects (HADS and LADS respectively). The observational
separation is not clear but typically {\it V}-band amplitude around 0.3~mag
is a borderline. HADS are more frequently radial mode pulsators with a few
modes observed and they are more evolved objects than LADS which have many
nonradial modes observed. We did not find any differences between high and
low amplitude objects. This may be a consequence of a low detection
threshold.

Fig.~6 shows the CMD for $\delta$~Sct stars superimposed on the diagram 
showing one OGLE-III subfield. 
\begin{figure}[htb]
\centerline{\includegraphics[width=12cm]{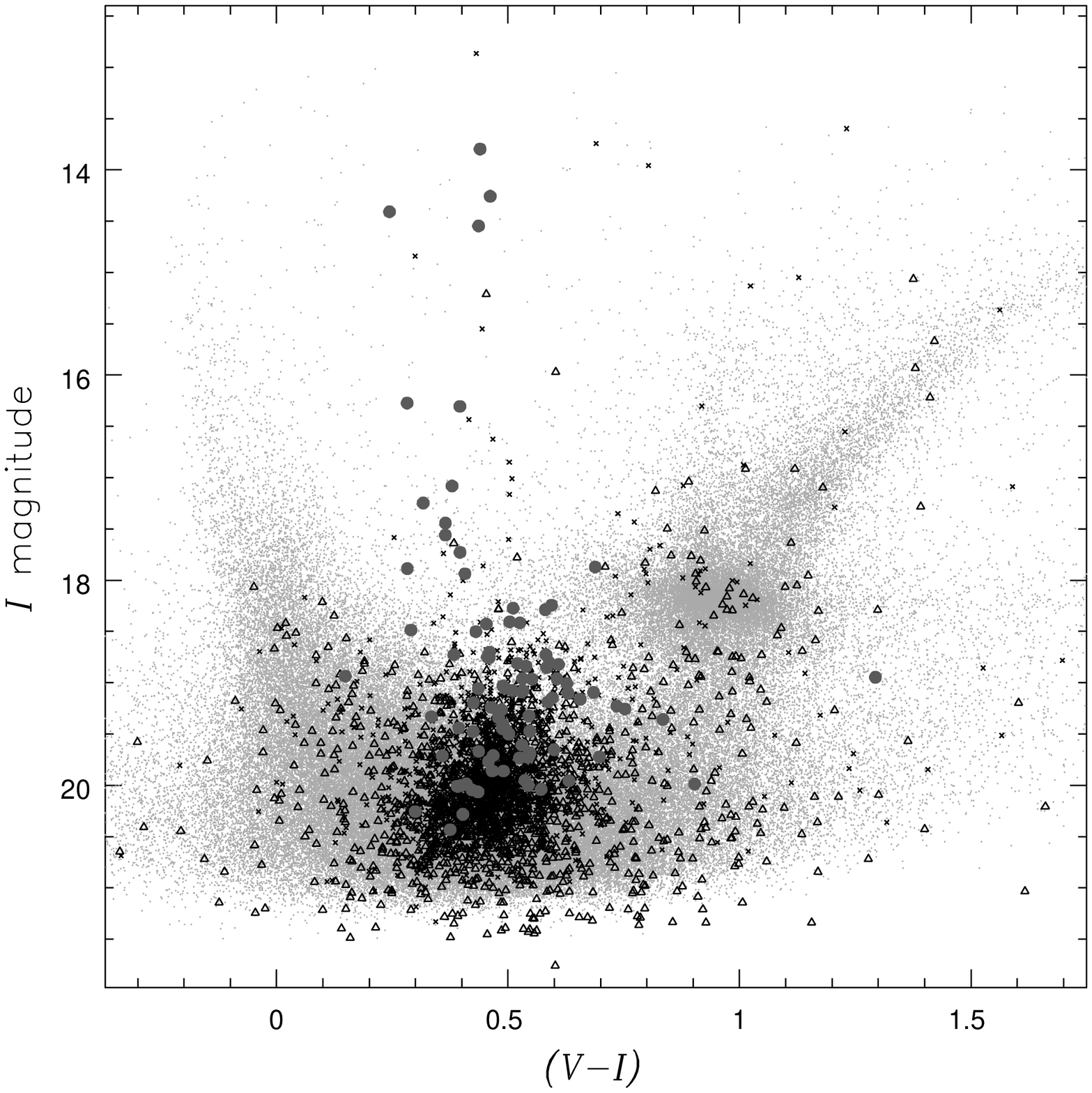}}
\FigCap{Color--magnitude diagram for $\delta$~Sct stars superimposed on the 
CMD of the OGLE-III subfield LMC100.1. Gray circles, crosses and open
triangles correspond respectively to multi-mode pulsators, uncertain
single-mode pulsators and the remaining single-mode pulsators.}
\end{figure}
Most of the stars have $(V-I)$ color between 0.3~mag and 0.7~mag. As the
stars are mostly faint ones ($I>19$~mag), the blending affects their
brightnesses and colors stronger than for \eg Cepheids.

The Fourier coefficients are frequently used parameters for describing 
the shape of the light curves of variable stars. We present calculated 
$\phi_{21}$ and $R_{21}$ parameters in  Fig.~7.
\begin{figure}[htb]
\centerline{\includegraphics[height=12cm, angle=270]{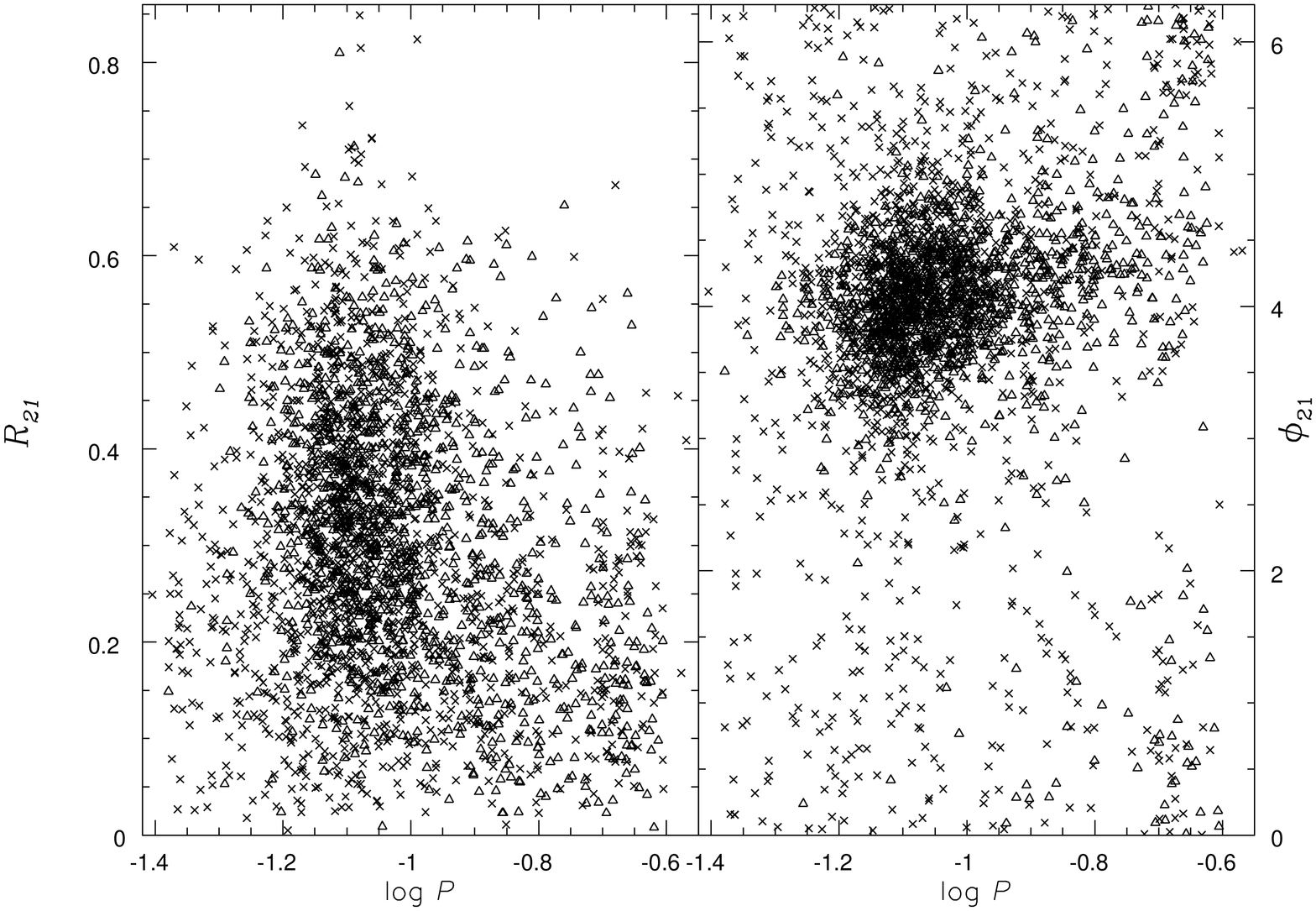}}
\FigCap{Fourier coefficients $R_{21}$ ({\it left panel}) and $\phi_{21}$ 
({\it right panel}) as a function of the logarithm of period. Uncertain 
objects are marked with crosses and all the other as open triangles.}
\end{figure}
The $R_{21}$ coefficient is typically bigger than 0.1 and smaller than 0.6.
For objects with period longer than around 0.12~days (\ie $\log P>-0.9$)
the values of this coefficient decrease. Most of the $\delta$~Sct stars
have $\phi_{21}$ close to 4. Objects with $\phi_{21}$ smaller than 3 or
larger than 5 are mostly marked as uncertain (crosses in Fig.~7). Similar
dependence of $R_{21}$ and $\phi_{21}$ on $\log P$ was found by other
authors (\eg Morgan \etal 1998, Poretti 2001). We did not find any
dependence of higher order coefficients on $\log P$.

The shortest period for sound single-mode $\delta$~Sct star in our catalog
is 0.041~d. It is much smaller than typical cadence of the OGLE-III
measurements in the LMC which is usually about 3~d. Sometimes the
observations of the given field are done more frequently but still very
rarely the cadence is smaller than the shortest periods discovered.
OGLE-III fields are overlapping, thus the observations of a star lying near
the CCD edge occasionally are obtained one after another. We checked a few
exemplary time-series using method described by Koen (2006). Although we
did not find Nyquist frequency to be smaller than any of the variability
frequencies of our objects we note the daily aliases are possible in some
cases.

\Section{Conclusions}
The most important result of the search for $\delta$~Sct stars in the LMC
is a high number of multi-mode pulsators found. The distance to this galaxy
is known rather accurately and together with measured luminosities gives
additional constrain for pulsation models.

Also a few Milky Way multi-mode $\delta$~Sct stars discovered here are of
interest. Up to eight frequencies were discovered and we can expect that
much larger number of such objects can be found in the OGLE-III Galactic
bulge data.

For the radial modes of the multi-mode objects we derived period-luminosity
relations. The scatter of points does not allow mode identification for
particular single-mode object. This scatter is also comparable to the
differences in luminosity for the F and the 1O modes at the same
period. Therefore, we conclude these relations cannot be used for mode
identification of single-mode $\delta$~Sct stars, even if absolute
luminosities are known.

Most of the multi-mode objects without the radial mode identification are
most probably members of the LADS subgroup of $\delta$~Sct stars.

More than a half of the presented objects are marked as uncertain.
Nevertheless, it is the biggest catalog of positively identified
$\delta$~Sct pulsators published so far in any environment including our
Galaxy.

\Acknow{Authors are grateful Prof.~W.~Dziembowski for fruitful
discussions. We thank Z.~Ko³aczkowski, A.~Schwarzenberg-Czerny and
J.~Skowron for providing the software used in this study. This work was
supported by the Foundation for Polish Science through the Homing (Powroty)
Program and by MNiSW grants: NN203293533 to IS and N20303032/4275 to AU.

The massive period search was performed at the Interdisciplinary Centre for
Mathematical and Computational Modeling of Warsaw University (ICM UW),
pro\-ject no. G32-3. We are grateful to Dr.~M.~Cytowski for helping us in
this analysis.}

\end{document}